# Formation of periodical metal oxide multilayer structures for the X-ray standing wave applications

**Authors:** Olga Dikaya[a]*, Dmitri Novikov[b], Alexander Goikhman[a], Ksenia Maksimova[a]

**Affiliations:** [a]Immanuel Kant Baltic Federal University, the Institute of Physics, Mathematics and Informatics, REC "Functional Nanomaterials", Gaidar st. 6, 236016 Kaliningrad, Russian Federation

[b]Deutsches Elektronen-Synchrotron DESY, Notkestraße 85, D-22607 Hamburg, Germany

*\*Corresponding author:* e-mail: odikaya@kantiana.ru, tel.: +7 4012 595595 ext. 9047, address: 236016 Gaidar st. 6, Kaliningrad, Russian Federation

**Abstract**

Periodical multilayer (ML) structures can be used as generators of X-ray standing waves (XSW) for investigation of objects and processes on solid/liquid and solid/gas interfaces. In this paper, we investigate the specific requirements to the structural properties of the multilayer structures for XSW application. We consider the effect of typical defects in the ML structure on the X-ray standing wave formation and show that the X-ray standing wave is very robust against the random imperfection in the multilayer structure. In contrast, the roughness of the topmost layer will have a strong influence on the XSW experimental results, as the ML serves as a support for the investigated objects, so that the surface geometry gets directly translated into the objects. In the experimental part of this work, we have used the ion-beam deposition to grow Ni/Al metal- and metal oxide-based multilayers and investigate with AFM their surface quality. The presented results demonstrate that metal oxides can be successfully used as basic material for X-ray multilayer standing wave generators.

**Abbreviations**

AFM – Atomic Force Microscopy

Rms – Root Mean Square

RF – Radio Frequency

XRR – X-Ray Reflectometry

XRD – X-Ray Diffraction

XSW – X-ray Standing Wave

XRF – X-Ray Fluorescence

PML – Periodical Multilayer

IBD – Ion-Beam Deposition

SEM – Scanning Electron Microscopy

## 1. Introduction

The periodical multilayer (PML) structures are widely used as optical elements for electromagnetic waves in a broad spectrum from the visible light to the hard X-ray radiation. In the X-ray range, their functionality is based on diffraction and allows combining monochromatization and focusing in a wide range of applications reaching from soft X-ray lithography to X-ray microscopy and astronomy. They are also routinely installed in X-ray spectrometers, diffractometers, at synchrotron beamlines, etc. [1], [2]. Control and characterization of the structural properties of periodical multilayers is also usually done by X-ray Bragg scattering [3]. Some less common applications of X-ray diffraction on PMLs lie beyond the areas of optics and multilayer diagnostics. One of them, the X-ray standing wave technique (XSW) [4], makes use of the periodic electromagnetic field formed by the interference of the incident and PML-diffracted X-rays. For a long time this technique was used for the high precision characterization of internal PML structure parameters, such as atomic impurities [5], interface roughness study [6], etc.

Meanwhile, the X-ray standing wave field also exists in the space above the multilayer surface and can give access to information about objects directly adjacent to the PML. The X-ray standing waves technique employing the PMLs as X-ray field generators is known since 1985 [7] and was used e.g. for the investigation of the metals deposited on the surface, ions at solid/liquid interfaces, absorption of organic molecules etc. In a general case, such XSW data provides element specific information about the spatial distribution of definite sorts of atoms in the object, including their distances to the PML surface and the width of atomic distributions. More details, e.g. the form of distribution, may be extracted if any *ab initio* information is available. The advantages of PML-generated XSW technique are obvious [8], [9]. First and foremost the period of the standing wave, which for the first order Bragg reflection coincides with the period of the periodic structure, can be matched to the characteristic dimensions of the object. Besides, the PML surface can be, without any losses for the experimental method, coated by a protective or chemically functionalized layers.

As the XSW experiment makes use of the fluorescent radiation emitted by the object, this imposes special requirements on the materials in the multilayer structure, as their own secondary X-ray emission may not overlay that of the target atoms. The same applies also to the impurities present in the bulk PML material and all contaminations that can be possibly introduced in the process of the multilayer production.

The structural perfection of the multilayer-generated field directly influences the quality of the experimental data. In the X-ray optics application, the usual figure of merit for the PMLs is the reflected beam intensity. It is achieved by a strict periodicity of the structure, avoiding intermixing and interface roughening. These features are also important for standing waves generators, as they ensure producing standing wave with high contrast. However, random growth failures and interface imperfections are less critical for the XSW experimental results, as the phase of the interference field stays largely unaffected by defects in the bulk of the multilayer. Meanwhile, the roughness of the surface, that has a minor influence on the Bragg reflectivity, starts to play a central role in the X-ray standing wave method: in case the PML serves as a physical support for the sample object, the surface roughness directly interferes with the distribution of atoms investigated in the experiment. In many practical cases, even a distortion due to interface roughness at the level of several angstrom is comparable with the intrinsic size of the object and complicates the data interpretation [10], [11], [12], [13], [14].

In this work we have used ion-beam deposition technique to produce periodic multilayer structures for X-ray standing wave generation. The multilayers are optimized for experiments on light ions at liquid interfaces and biological samples. This defines the choice of materials, which will allow good PML structure control, low surface roughness and at the same time have no own X-ray emission lines that can contaminate the target sample emission in the range from 1.5 keV to 6 keV. We investigated the substance combinations Ni/Al, $AlO_x/NiO_x$ and $CuO_x/AlO_x$ on $Si/SiO_2$ and $Al_2O_3$ substrates and analyze experimentally the interfacial and surface roughness of the PMLs. We also give a theoretical analysis of the influence of multilayer imperfections on the generated X-ray standing wave parameters.

The rest of this paper organized as follows: Section 2 "Experimental methods" presents used experimental techniques. In Section 3 "Periodic multilayer growth and characterization", we present detailed investigation of grown metal and oxide PML structures. In Section 4 "Growth errors effects: model XSW calculations", we demonstrate the influence of discovered errors in PML growth on the formation of X-ray standing wave. Finally, in Section 4 "Conclusions" we summarize all results of this paper.

## 2. Experimental methods

### 2.1. Deposition technique.

All PML samples are deposited with an ion-beam sputtering system based on a 12cm radiofrequency (RF) Kaufman ion source with RF neutralizer. We used Ar or Kr to generate 0.5-1.5 KeV ion beam, and $O_2$ gas (purity 99.999%) with precise flow control (0.03 – 3 l/s) for the oxidation processes [15]. The three-position rotating target holder was used to deposit all layers in one vacuum process (Fig.1. a) from high purity targets of Al (99.9%), Ni (99.99%), and Cu (99.9%).

To achieve single layer uniformity we use the planetary substrates holder, which rotates the substrates in the two paths simultaneously (Fig.1. b). The deposition process starts with the basic pressure in the vacuum chamber $10^{-5}$ Pa, and proceeds with an Ar or Kr pressure $2.4*10^{-2}$ Pa, or rises up to $4.2*10^{-2}$ Pa in case of oxide growth. A quartz sensor was used for *in situ* control of the layer thickness (Fig.1. a).

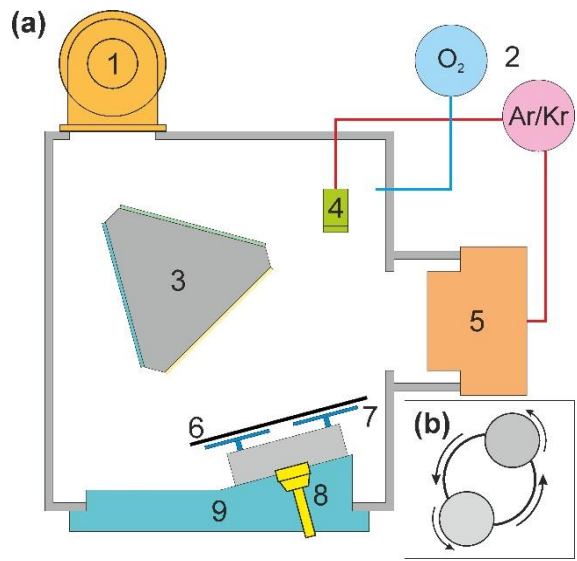

**Figure 1.** Ion-beam deposition system. (**a**) Vacuum chamber: 1 – cryopump, 2 – gas system, 3 – targets holder, 4 – RF neutralizer, 5 – RF source, 6 – shutter, 7 – planetary substrate holder, 8 – quartz control system, 9 – vacuum chamber door. (**b**) Rotation trajectory of planetary holder.

In order to avoid the temperature diffusion on the interfaces of the PML layers [16], the Ni/Al multilayer are deposited with a time interval between the layers. For the same reason, the metallic materials are deposited with the 500 eV ion beam. The oxide multilayers are deposited without time intervals by the 1 keV ion beam. In addition, the $CuO_x/AlO_x$ PML are deposited with thicker period in case of surface and interface roughness tests. Also, copper is one more promising material for XSW applications in biological area. All samples are grown at room temperature on the naturally oxidized Si (100) substrates.

### 2.2. Characterization instruments.

All samples were examined by the atomic force microscopy (Aist SmartSPM-1000) in the tapping mode using a NSG30 cantilever with the curvature radius 10 nm. Additionally, the metal oxide samples were investigated by the scanning electron microscopy (SEM) using Carl ZEISS Crossbeam 540 setup with the accelerating voltage 10 kV.

The multilayer samples were characterized by the XRR technique using a laboratory X-ray diffractometer Bruker D8 Discover with the Cu-Kα source (λ=0.154 nm) and were fitted with the DIFFRAC.LEPTOS (commercial software from the Bruker company) by the genetic algorithm instruments. We used the θ/2θ grazing incidence angle geometry for the experiment with the θ range [0.2°; 2.0°].

**3. Periodic multilayer growth and characterization.**

During the growth experiment we deposited two periodical multilayer structures with nominal thickness 120 nm corresponding to 20 bilayers with 6 nm thickness. The growth parameters was chosen so that the thickness of each individual layer is equal to 3 nm. As will be seen below, some material interaction issues may have influence on final thickness ratio. As was mentioned in paragraph 2.1, Ni and Al have great depth of interpenetration [16], so that characteristic of materials may cause formation of the transition layer which formally changes ratio of thicknesses.

As the first step of growth experiment, the thin films of Ni, Al, $NiO_x$, $AlO_x$ were deposited on silicon substrates with the surface roughness 3.26 Å and waviness (long range Rms) 9.56 Å. AFM investigations of 2x2 μm area of the 3 nm thick single layer samples provided basic information about the local surface and long range roughness formation (Table I).

**Table I.** The surface roughness of the single-layer samples.

|  | Ni | Al | $NiO_x$ | $AlO_x$ |
|---|---|---|---|---|
| Local Rms | 6.1 ± 0.6 Å | 6.6 ± 0.6 Å | 1.4 ± 0.2 Å | 1.5 ± 0.2 Å |
| 2x2 μm Rms | N/A | 12.7 ± 2 Å | 9.34 ± 1.4 Å | 7.3 ± 1.5 Å |

The decrease of oxide thin film roughness relative to that for the metallic films is explained by the different growth mechanism at a $SiO_2$ surface. While aluminum and nickel form 3D metal islands at the initial stage of the film nucleation, oxides of these metals start to grow with formation of condensation centers, uniformly distributed over the surface, and proceed to a layer-by-layer mode developing smooth thin film without distinct grains.

Periodic Ni/Al, AlO$_x$/NiO$_x$, CuO$_x$/AlO$_x$ multilayers were grown on Si/SiO$_2$ substrates. The top layer (Al, NiO$_x$, AlO$_x$) morphology of each PMLs type is demonstrated on the Fig.2. Surface of Ni/Al consists of grains with a typical lateral size of about 40 nm. The root mean square roughness values measured with AFM are shown in Table II. Remarkably, roughness of the metallic multilayer is lower than that of separate single layers. This is explained by the roughness compensation inherent for the growth mechanism [22]. The oxide based PMLs show substantially lower surface roughness values, which will facilitate better results in the XSW experiment data interpretation.

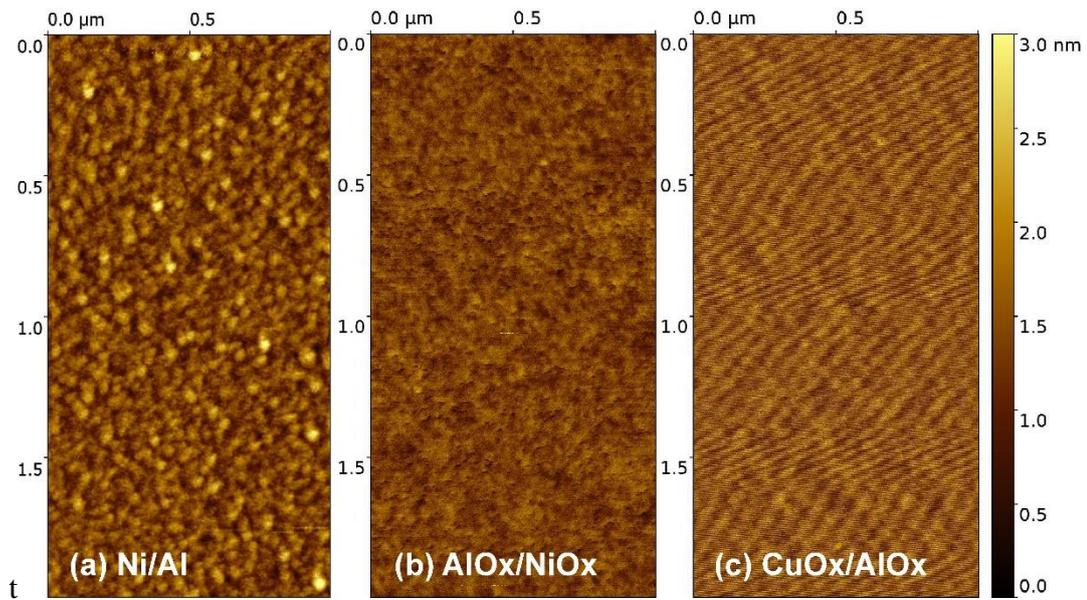

**Figure 2.** The atomic force microscopy surface images of multilayers **(a)** Al oxidized in the atmosphere, **(b)** NiO$_x$, and **(c)** AlO$_x$ top layers.

**Table II.** The surface roughness of the multilayer samples.

|  | Ni/Al | AlOx/NiOx | CuOx/AlOx |
|---|---|---|---|
| Local Rms | 3.77 ± 0.6 Å | 2.42 ± 0.3 Å | 2.71 ± 0.5 Å |
| 2x2 μm Rms | 13.76 ± 3.8 Å | 10.76 ± 2.4 Å | 9.77 ± 2.7 Å |

Typical SEM images from the cross-section AlO$_x$/NiO$_x$ and CuO$_x$/AlO$_x$ multilayers shown on the Fig. 3 (a, b) prove an excellent uniformity of the multilayers. The experimental XRR curves (Fig. 4, Table III) show good reflectivity of the first order Bragg reflection and a clear Kiessig fringes structure, typical for clean and well-defined interfaces.

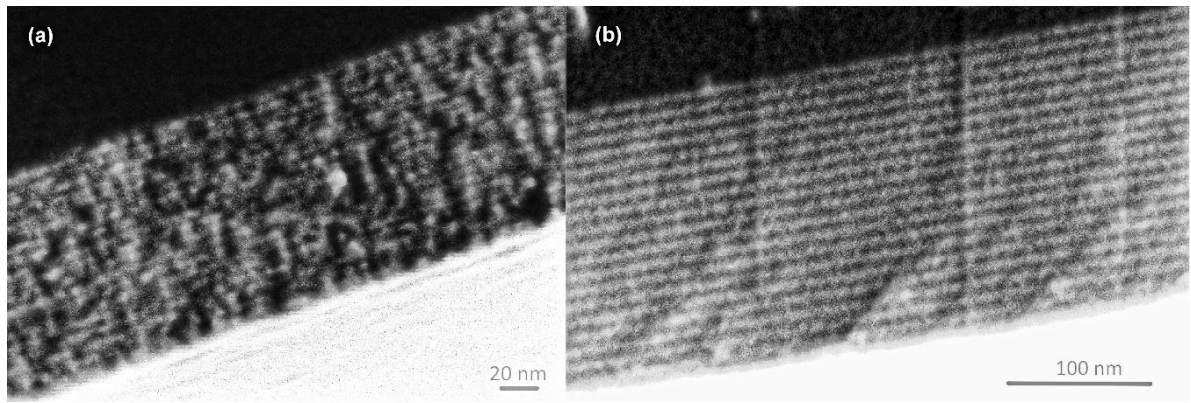

**Figure 3.** The scanning electron micrographs in high resolution mode of **(a)** AlO$_x$/NiO$_x$ and **(b)** CuO$_x$/AlO$_x$ PMLs.

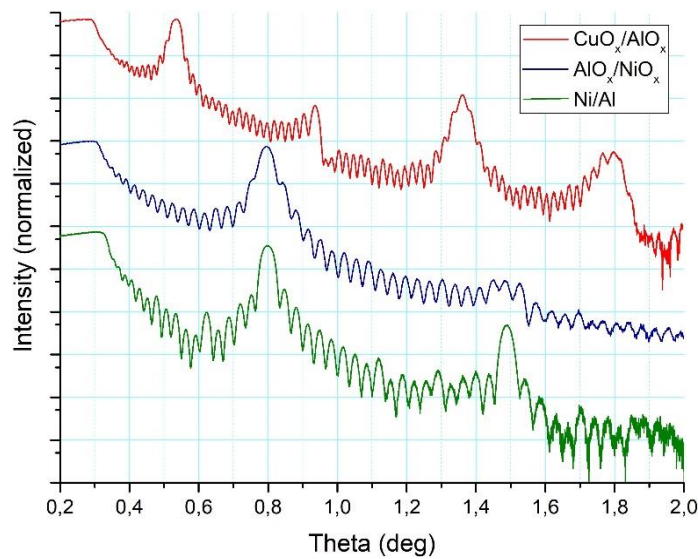

**Figure 4.** The reflectivity curves of the Ni/Al, AlO$_x$/NiO$_x$, and CuO$_x$/AlO$_x$ multilayers obtained with the Bruker D8 Discover setup (Cu-K$\alpha$). All curves normalized relative to the critical angle (maximum reflectivity at the critical angle is equal to 100%), curves corresponding to AlO$_x$/NiO$_x$, and CuO$_x$/AlO$_x$ are shifted linearly for more clear representation.

**Table III.** The XRR fitting results from the structures.

| Structure | Period, nm | Interface Rms, Å | Reflectivity, 1st Bragg peak, % |
| --- | --- | --- | --- |
| Ni/Al | 3.23 ± 0.3 / 2.77 ± 0.45 | 6.46 ± 0.6 / 14.52 ± 4.1 | 50 |
| $AlO_x$/$NiO_x$ | 2.95 ± 0.2 / 3.05 ± 0.17 | 6.2 ± 0.9 / 5.8 ± 0.12 | 43 |
| $CuO_x$/$AlO_x$ | 4.8/5.2 ±0.2 | 2 ± 0.1 | 68 |

As one can see from the Table III, the Ni/Al multilayer has a high divergence from mean layers thicknesses and very rough interfaces, especially at the interface of Ni layers grown on Al. These data correlate with surface study of the single layers by the AFM showing higher roughness of the Al layer than of the Ni layer. In closer inspection, roughness reduces through the structure from bottom (substrate) to top in accordance to calculations by *Goray et. al.* [22].

The results show that the $AlO_x$/$NiO_x$, and $CuO_x$/$AlO_x$ are well suited for application in X-ray standing wave generators. The list of basic substances can be obviously expanded according to the specific needs of the experimental requirements. A further reduction of the top layer roughness might be reached by use of specially polished substrates and introduction of buffer layers for strain compensation. Finally, the additional top layer of "smoothing" material may be deposited or the surface of PML may be reconstructed by the ion etching.

**4. Growth errors effects: model XSW calculations.**

It should be noted that all simulations of the X-ray standing wave field in this paper were performed with the TER_sl software package [17] at the X-ray server by *S. Stepanov* [18].

The general requirements to the multilayer structure quality include low interface and surface roughness and the high uniformity of the layer thickness. The strict periodicity of the layer thickness is the key parameter for the classical X-ray mirrors. In practice, physical deposition methods may provide minor thickness errors dependent upon several factors: vacuum chamber construction, deposition rates, type of thickness control system etc. As was pointed out, e.g. by *Spiller et. al.* [19], while the thickness errors reduce the reflectivity for all peaks, the higher reflection orders are stronger affected than the 1st Bragg peak.

As the X-ray standing wave experiments usually use the first order Bragg reflection, the requirements to periodicity get therefore less stringent. Still, any major errors may cause a deformation of the 1st Bragg peak standing wave and must be avoided. In the literature it has been proved for the depth-graded multilayer systems [20] that random minor growth errors has negligible influence on the 1st Bragg peak shape and its reflectivity. In contrast, systematic minor errors dramatically change the 1st Bragg peak and reduce its reflectivity. For the PMLs grown for standing wave generation, these factors are less important. In contrast, a single major thickness error in a layer may cause the displacement of all the subsequent reflecting layers, which effectively splits the PML into two Bragg reflectors with two different standing waves produced in one structure.

Let us consider an example with a single significant error in the structure, e.g. a 5 Å "shift layer" between the periods in a PML with 6 nm periodicity. The comparison of the curves corresponding to the different "shift layer" positions are presented on the Fig. 5. One can see that the shape of the reflectivity curve strongly depends on the error location inside the PML layer stack. Remarkably, in all cases the 1st Bragg peak of such structures has reasonably small divergence in comparison to the ideal structure. However, for larger reflection orders one can observe notable shift and splitting of the Bragg peaks. The influence of a single growth error on the X-ray standing wave is shown at Fig. 5b. Interestingly, despite the shift of ~10% between two sets of multilayers that generate the XSW, the overall form of resulting field demonstrates only a slight variance for different structures. However, the phase of the XSW appears to be sensitive to the growth error of that sort.

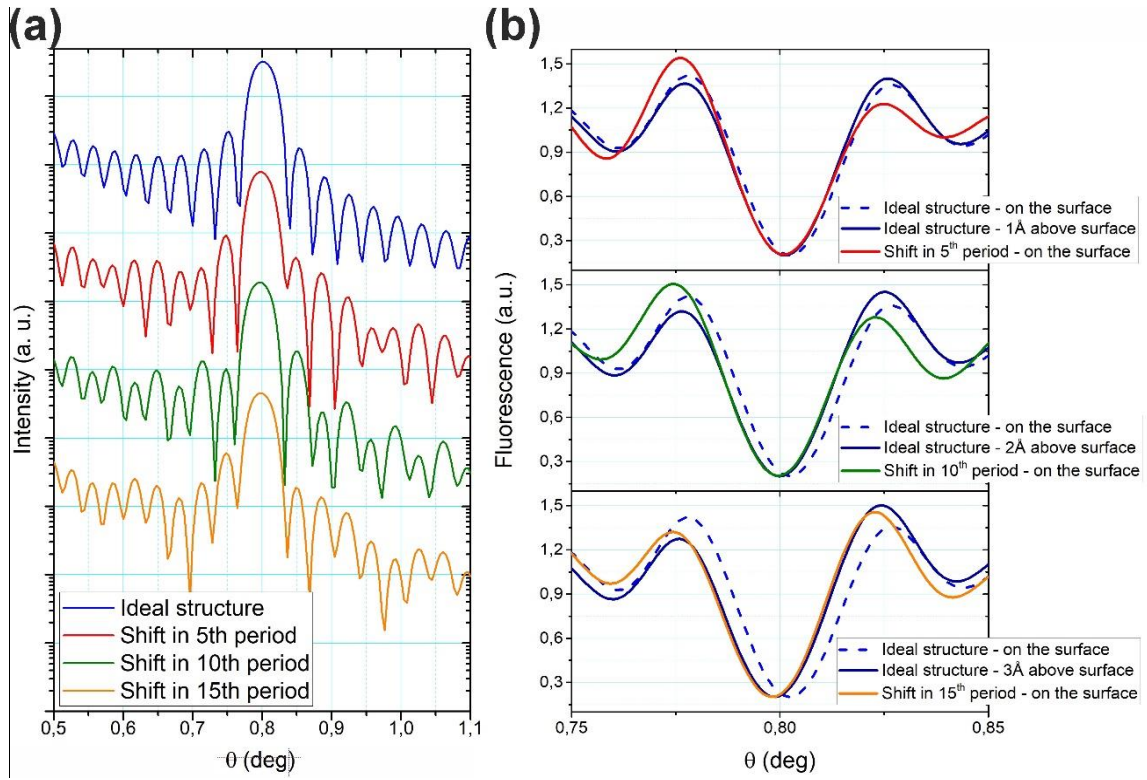

**Figure 5.** Simulated reflectivity and fluorescence curves (Cu-Kα λ=0.154 nm) for the structure NiO$_x$/AlO$_x$ with 20 periods, bilayer thickness 6 nm, interface roughness 2 Å, and an additional 0.5 nm "shift layer" at different positions inside the layer stack (periods counted from the substrate to the surface of the structure). **(a)** The shift layer influence on 1$^{st}$ Bragg peak form. **(b)** The XSW phase at the surface of the PML with a "shift layer". Along with the curves the ideal PML and structures with "shift layer", calculated for the multilayer stack surface, we also show an ideal PML curve for an ideal PML that would be most similar in form and phase to that for the non-ideal PML. The displacement value for the latter curve gives an estimate for the potential error in data evaluation that could arise due to the usage of PMLs with a single layer thickness mistake.

Effectively, the "shift layer" splits the PML into two sections, each of them producing its own standing wave. At the surface, the correct XSW generated by upper section is overlaid by the shifted XSW of the by the underlying part. Remarkably, the influence of the "shift layer" depends on its position inside the multilayer stack. If the shift layer is positioned deeper in the stack, the XSW is dominated by the well pronounced field formed in the upper PML section and has detectable phase shift at the surface. In the opposite case, the substantial part of the XSW is formed underneath the shift layer and the physical surface of the stack is displaced relative to the designed standing wave phase.

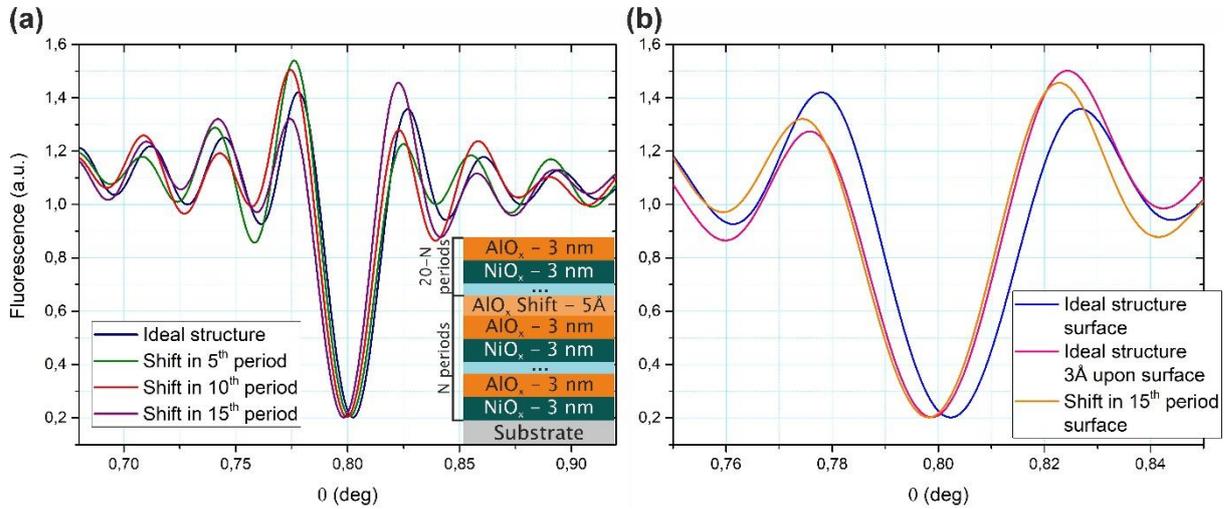

**Figure 6.** Simulated XSW curves (10 Å above the surface) for the structure $NiO_x/AlO_x$ with 20 periods, bilayer thickness 6 nm, interface roughness 2 Å with the shift 5 Å: no shift, a shift in the 5th period, in the 10th period and in the 15th period. The periods are counted from the substrate to the surface of the structure. The changes in the standing wave phase introduced by the growth error will directly influence the data evaluation and provide wrong values for the distance of object atoms to the PML surface.

The inner interface roughness is a key parameter for the optimum reflectivity of PMLs [19]. *Kawamura et. al.* [21] have modelled a variation of the XSW curves for the different interface roughness and demonstrated that the intensity contrast of the XSW curve gets smoothed with interface roughness ascending.

The influence of the surface roughness is less obvious. On the one hand, it has a minor influence on the reflectivity and the 1st Bragg peak shape. As one can see in Fig. 7, the X-ray standing wave phase is practically independent of the surface roughness. It is clear that the XSW intensity decrease is negligible for a practical applications: the difference between the curves with a minimum and maximum roughness is only 0.05% from total amplitude of the curve, while the phase stays unaffected.

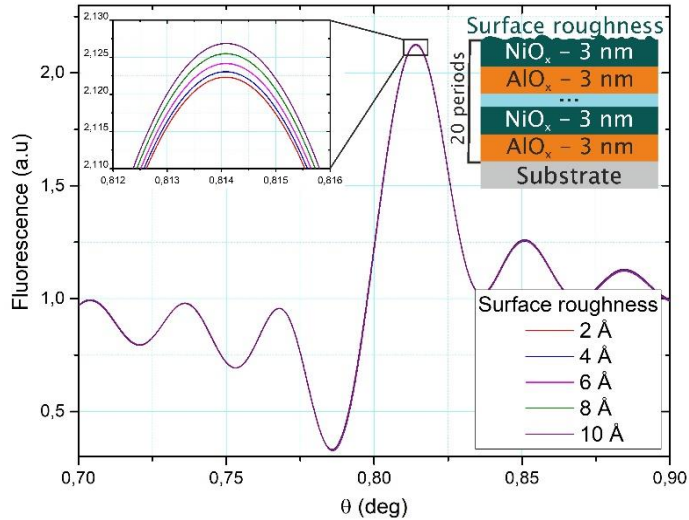

**Figure 7.** Simulated XSW curves for a single atomic layer of object atoms positioned 15 Å above the surface as a function of surface roughness for the structure $AlO_x/NiO_x$ with 20 periods, period 6nm and the interface roughness 2 Å. As predicted by the theory of X-ray scattering, the surface roughness has no measurable effect on the phase of the XSW field and manifests itself in a minor influence on the XSW amplitude.

However, in the cases when the PML serves as a support to the investigated objects, an uneven surface will smear the XSW values for the object atom distribution to the extent controlled by the roughness of the top layer. In practice, the impact of the surface geometry influence depends also on the characteristic dimensions of the objects and the PML period. Anyway, reducing the surface roughness becomes the primary goal one must achieve in the growth process of PMLs that will be used in the interface studies.

## 4. Conclusion

In conclusion, we have investigated the possibilities to use Ni/Al, AlOx/NiOx, and CuOx/AlOx periodic multilayer structures as generators of X-ray standing wave. Model calculations have shown, that the X-ray standing wave formed by the first Bragg reflection order is very robust against minor random errors in the periodic multilayer. The key parameter of the multilayer structures for the XSW applications is the surface roughness, and all efforts should be focused on the reduction of this factor.

The multilayers were produced by the ion-beam deposition technique with Kr- and Ar-plasma and investigated by AFM and X-ray diffraction. The surface roughness in PMLs produced by the IBD method appears to be largely independent from the roughness of the underlying interfaces and is mainly controlled by the properties of the top single layer. In this work, we have reached the Rms roughness values of ~ 3 Å for the $AlO_x$ surface. A further improvement can be achieved by changing the top layer for e.g. the $CuO_x$, which would though require a change in the XSW sample preparation.

Functional characterization showed a substantial difference in the quality of the metallic Ni/Al and metal oxide structures: the $AlO_x/NiO_x$ and $CuO_x/AlO_x$ demonstrate lower interface roughness, and the XRR analysis also revealed less layer thickness error. This suggests that oxides should be preferred as basic materials to insure high reflectivity. A comparison of $AlO_x/NiO_x$ and $CuO_x/AlO_x$ structures shows, that the latter combination has better interface properties and provides an X-ray standing wave with a better contrast. The ion-beam deposition technology allows using multiple target materials and is carried out in one vacuum cycle, thus preventing any contamination of the multilayers. One should also mention the very moderate manufacturing costs, which allows a large flexibility in customizing of PML parameters to match specific experimental conditions.

**Acknowledgements**